\documentclass[aps,pra,twocolumn,amsmath,amssymb,showpacs]{revtex4}

\newcommand{\ket}[1]{|{#1}\rangle}

\usepackage[dvips]{graphicx}
\usepackage{mathrsfs}

\begin{document}

\title{Quantum gate characterization in an extended Hilbert space}

\author{Peter P. Rohde}
\email[]{rohde@physics.uq.edu.au}
\homepage{http://www.physics.uq.edu.au/people/rohde/}
\author{G. J. Pryde}
\author{J. L. O'Brien}
\author{Timothy C. Ralph}
\affiliation{Centre for Quantum Computer Technology, Department of Physics\\ University of Queensland, Brisbane, QLD 4072, Australia}

\date{\today}

\begin{abstract}
We describe an approach for characterizing the process performed by a quantum gate using quantum process tomography, by first modeling the gate in an extended Hilbert space, which includes non-qubit degrees of freedom. To prevent unphysical processes from being predicted, present quantum process tomography procedures incorporate mathematical constraints, which make no assumptions as to the actual physical nature of the system being described. By contrast, the procedure presented here assumes a particular class of physical processes, and enforces physicality by fitting the data to this model. This allows quantum process tomography to be performed using a smaller experimental data set, and produces parameters with a direct physical interpretation. The approach is demonstrated by example of mode-matching in an all-optical controlled-\textsc{NOT} gate. The techniques described are general and could be applied to other optical circuits or quantum computing architectures.
\end{abstract}


\maketitle

Quantum information science promises information processing and transmission capabilities far beyond what is achievable using classical physics. In particular, quantum computing has the potential to solve problems which are intractable on classical computers. Processing quantum information requires quantum gates designed to implement unitary transformations on up to a few qubits. In practice experimental quantum gates perform a process which approximates the desired unitary operation and usually includes some decoherence. Characterizing these quantum processes is critical. This can be achieved through quantum process tomography (QPT) \cite{bib:ChuangNielsen97,bib:Poyatos97,bib:NielsenChuang00,bib:White03,bib:Gilchrist04,bib:Pryde04}, which expresses the experimental process in terms of a basis of unitary operations which span the space of allowed operations. Experimental noise usually results in unphysical process reconstructions and a maximum-likelihood correction procedure \cite{bib:Pryde04,bib:Ziman05} must be used to find the nearest physical process. An alternative approach, which we present here, is to construct a physical model of the experimental gate by extending the Hilbert space to include non-qubit degrees of freedom. Experimental data is then fitted to this model to infer the parameters describing the system. This step replaces the standard maximum-likelihood correction procedure. The quantum process in the qubit space is inferred by tracing out these additional degrees of freedom from the model, whereupon mixing and decoherence effects manifest themselves.

A quantum computer is a large interferometer \cite{bib:Ekert04}, and it is likely to be mode-mismatch of interfering modes that ultimately limits performance. We are therefore motivated to model imperfections in gate performance in terms of mode-mismatch. This requires us to consider the non-qubit degrees of freedom of physical qubits. This approach is most natural for optical gate implementations, whereby the spatio-temporal structure of photonic qubits must be considered, however, in principle it could be applied to quantum gates in any physical architecture.

Here we consider linear optics quantum computing (LOQC) \cite{bib:KLM01} in particular. We construct a model which explicitly allows for the effects of mode-mismatch, whereby photon indistinguishability is compromised within a circuit, thereby undermining interference effects. We show that by fitting the parameters in the model to experimental data, the mode-matching characteristics of an experimental gate can be obtained. From the model, simulated measurement probabilities can be determined analytically, which are immune from the effects of experimental noise. This allows QPT to be performed without requiring maximum-likelihood correction to ensure physicality. This procedure could be applied to other quantum computing architectures by identifying the physical processes of importance and constructing a suitable gate model. Importantly, this technique reduces the range of input states and measurement bases required to reconstruct the process, which may be of considerable advantage for architectures where the full range of measurements is not possible.

\begin{figure}[!htb]
\includegraphics[width=0.5\textwidth]{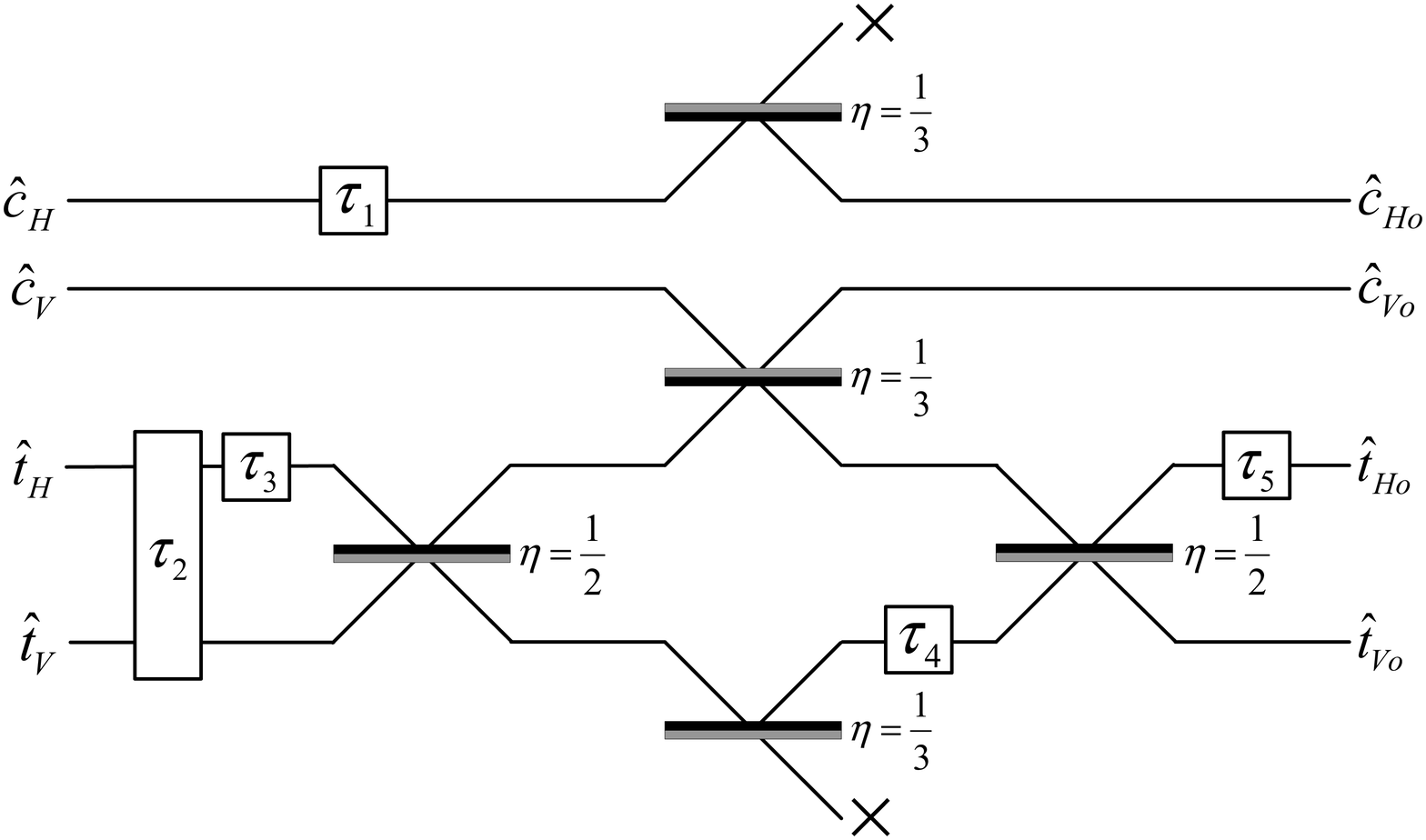}
\caption{\label{fig:cnot_schem}
Schematic of the \textsc{CNOT} gate using beamsplitters with reflectivities $\eta$. $c$ and $t$ denote the control and target qubits respectively. Modes labeled `$\times$' are discarded and serve to balance the amplitudes in the different paths. The gate is post-selected upon detection of exactly one photon between the $c$-modes and one between the $t$-modes. $\tau$-boxes represent mode-mismatch parameters, described in the text. We adopt the phase-asymmetric beamsplitter convention, whereby sign-inversion takes place upon reflection from the \emph{gray} beamsplitter surfaces.}
\end{figure}

Several in-principle demonstrations of elementary LOQC gates have recently been performed \cite{bib:Pittman01,bib:OBrien03,bib:Zhao05,bib:Gasparoni04}. We illustrate our techniques by example of an LOQC implementation of the \textsc{CNOT} gate \cite{bib:Ralph02}, shown in Figure \ref{fig:cnot_schem}. The gate employs dual-rail logic whereby a qubit is encoded across two spatial modes of a single photon. The gate is non-deterministic and post-selected upon detection of exactly one photon across the control modes and one across the target modes. By considering a gate which operates using coincidence detection we significantly simplify the decoherence effects we need to consider. However our approach can be also used to analyze heralded gates \cite{bib:RohdeRalph05,bib:Ralph01}. Our experimental gate construction is identical to that reported in \cite{bib:OBrien03}. Experimentally, beamsplitters are implemented using waveplates and polarizing beam displacers, allowing splitting ratios and phase delays to be set with a high degree of accuracy. We therefore restrict ourselves to considering the effects of mode-mismatch.

We introduce a representation for photons which explicitly captures their spatio-temporal structure. Specifically, we represent photons in terms of their wave-function across the photon degrees of freedom (\mbox{\emph{e.g.}} space, time, polarization \emph{etc.}),
\begin{equation}
\ket{\psi}=\int_{k_1}\!\!\!\dots\!\int_{k_n}\psi(k_1,\dots,k_n)\hat{\textbf{a}}^{\dag}_{k_1,\dots,k_n}\mathrm{d}k_1\dots \mathrm{d}k_n\ket{\textbf{0}}
\end{equation}
where \mbox{$\psi(k_1,\dots,k_n)$} is the joint wave-function across the photon degrees of freedom \mbox{$k_1\dots k_n$}, and \mbox{$\hat{\textbf{a}}^{\dag}_{k_1,\dots,k_n}$} is the single photon creation operator at the corresponding infinitesimal position, time \emph{etc.} This is a more generalized version of the representation adopted in \cite{bib:RohdeRalph05} when studying input distinguishability effects in LOQC.

The \textsc{CNOT} gate described strictly requires complete path indistinguishability at all locations in the circuit where photonic interactions take place. When photon indistinguishability is compromised non-ideal gate operation ensues and the gate no longer implements the \textsc{CNOT} logical transformation. The introduction of such distinguishability is generically referred to as \emph{mode-mismatch} and is one of the major challenges facing experimental implementations. Mode-mismatch may arise for a number of reasons: imperfect spatial overlap between photons; imperfect temporal synchronization; differing center frequencies or bandwidths; differing polarization; or, any other effect which introduces distinguishing information between photons. These problems manifest themselves not only for separate photons, but also for single photons where self-interference takes place between different paths.

We model mode-mismatch by introducing displacements (\mbox{\emph{i.e.}} distinguishability) into the photon wave-function at different points in the circuit. Figure \ref{fig:cnot_schem} shows the five locations where displacements are introduced, labeled \mbox{$\tau_1\dots\tau_5$}. These displacements are sufficient to model arbitrary mode-mismatch effects (\mbox{\emph{i.e.}} the introduction of additional displacements will be redundant). When a photon passes through a $\tau$-box it undergoes the transformation
\begin{equation}
\psi(k_1,\dots,k_n)\to\psi(k_1+\tau_{m,1},\dots,k_n+\tau_{m,n})
\end{equation}
where $\tau_{m,n}$ is the displacement introduced at location $m$ into the $n^\mathrm{th}$ photon degree of freedom. We have assumed $\psi(k_1,\dots,k_n)$ to be Gaussian for simplicity. Arbitrary forms for the wave-function could be chosen, however, specifically using Gaussians does not affect the generality of the model or detract from its predictive power. This is because it is the degree of wave-function overlap which is of significance, and, regardless of its form, there will always be a set of $\tau$'s corresponding to a given degree of overlap. However, depending on the form and variance of the wave-function, the magnitude of the $\tau$ parameters corresponding to a given degree of distinguishability will change. This complicates a quantitative interpretation of the parameters. Instead they are best interpreted in terms of their relative magnitude.

In addition to mode-mismatch, the model accommodates for distinguishability which arises during state preparation (implicitly incorporated into $\tau_1$, $\tau_2$ and $\tau_3$) and measurement ($\tau_1$ and $\tau_5$). Parameters $\tau_1$ and $\tau_5$ do not affect circuit operation when operating in the computational basis. This is because state preparation (measurement) in a non-computational basis is equivalent to the introduction of beamsplitters before (after) the circuit to generate the required superposition. When operating in the computational basis these \emph{virtual} beamsplitters are not used and therefore do not form interferometers, rendering terms $\tau_1$ and $\tau_5$ irrelevant. Other optical circuits could be modeled in a similar way by identifying the locations in the circuit where mode-mismatch could occur and constructing an appropriate circuit model.

It is intuitive that, in the context of gate operation, mode-mismatch in any single degree of freedom is completely equivalent to mode-mismatch in any other single degree of freedom. This is because it is the magnitude of photon distinguishability, not the degree of freedom in which it is introduced, which results in non-ideal gate operation. However, it is not obvious that mode-mismatch which occurs in multiple degrees of freedom is equivalent to mode-mismatch in a single degree of freedom. We now demonstrate that this is the case, and consequently a single degree of freedom in photon distinguishability is sufficient to model arbitrary mode-mismatch effects. 

\begin{figure}[!htb]
\includegraphics[width=0.5\textwidth]{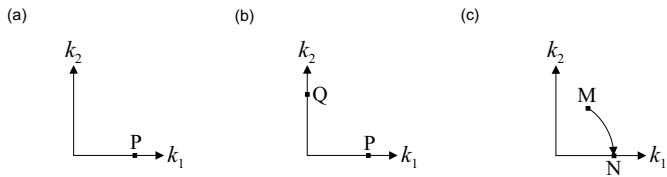}
\caption{\label{fig:parm_space}Graphical representation of the mode-mismatch parameter space, in two degrees of freedom only \mbox{($k_1$ and $k_2$)}. (a) Mode-mismatch in a single degree of freedom, represented by the point $\textbf{P}$. (b) Equivalence of mode-mismatch in any single degree of freedom, represented by the points $\textbf{P}$ and $\textbf{Q}$. (c) Equivalence of mode-mismatch in multiple of degrees of freedom, $\textbf{M}$, and a single degree of freedom, $\textbf{N}$, through rotational invariance in the choice of axes.}
\end{figure}

We introduce a geometric representation for the mode-mismatch parameter space, as shown in Figure \ref{fig:parm_space}. We let each axis of the graph represent a particular photon degree of freedom, and every point on an axis a vector quantity, representing the mode-mismatch parameters in that degree of freedom. From the point of view of gate operation, a point on the $k_1$-axis is completely equivalent to the corresponding point on the $k_2$-axis. If mode-mismatch occurs in multiple degrees of freedom, the contribution from the different degrees of freedom results in a point away from the main axes. However, the choice of axes for photon degrees of freedom are completely arbitrary and therefore the set of equivalent mode-mismatch parameters must be rotationally invariant in the space of photon degrees of freedom. Thus, any point in mode-mismatch parameter space can be rotated onto one of the main axes, \mbox{\emph{i.e.}} into a single degree of freedom. Therefore it is sufficient to express photons as a weighted integral over a single degree of freedom. Equivalently, five scalar quantities are sufficient to completely characterize the gate's mode-mismatch.  This geometric argument generalizes to higher dimensions.

The inherent equivalence of the mode-mismatch degrees of freedom unfortunately raises obstacles in the experimental interpretation of the parameters. Specifically, upon inspection of the parameters it is not possible, in principle, to determine in which photon degree(s) of freedom the mode-mismatch is occurring. For example, if temporal mismatch occurs this will manifest itself in exactly the same way as spatial mismatch and there is no way from the parameters to infer which is taking place.

Using the gate model we are able to derive analytic expressions for arbitrary coincidence measurement expectation values given arbitrary input states. We construct an 8 by 8 matrix $\textbf{M}_\mathrm{exp}$, of experimentally determined coincidence expectation values, and a corresponding matrix $\textbf{M}_\mathrm{model}$, of analytic expressions derived from the gate model. The rows of the matrices correspond to the input states $\ket{00},\ket{01},\ket{10},\ket{11},\ket{++},\ket{+-},\ket{-+},\ket{--}$, and the columns to the corresponding measurement settings, where $\ket{\pm}=\frac{1}{\sqrt{2}}(\ket{0}\pm\ket{1})$. We define the error matrix as
\begin{equation}
\textbf{M}_\mathrm{error}=|\textbf{M}_\mathrm{exp}-\textbf{M}_\mathrm{model}|
\end{equation}
where the absolute value is performed element-wise. From $\textbf{M}_\mathrm{error}$ we define the maximum and mean errors
\begin{eqnarray}
\mathscr{E}_\mathrm{max}&=&\mathrm{max}(\textbf{M}_\mathrm{error})\nonumber\\
\mathscr{E}_\mathrm{mean}&=&\mathrm{mean}(\textbf{M}_\mathrm{error})
\end{eqnarray}
We also consider the process fidelity $F_P$ \cite{bib:Raginsky01,bib:Gilchrist04,bib:Pryde04}, defined as
\begin{equation}
F_P=\mathrm{tr}\Big(\sqrt{{\chi_A}^{1/2}{\chi_B}{\chi_A}^{1/2}}\Big)^2
\end{equation}
where $\chi_A$ and $\chi_B$ are the process matrices of the processes being compared.

We minimize $\mathscr{E}_\mathrm{max}$ by optimizing across the mode-mismatch parameters \mbox{$\tau_1\dots\tau_5$}, which we label $\tilde{\boldsymbol{\tau}}_\mathrm{min}$. These results are substituted back into the gate model to generate an optimized model. The parameters $\tilde{\boldsymbol{\tau}}_\mathrm{min}$ are tabulated in Table \ref{tab:taus}.
\begin{table}[!ht]
\begin{tabular}{|c|c|c|c|c|c|}
\hline
Parameter	&	$\tau_1$		&	$\tau_2$		&	$\tau_3$		&	$\tau_4$		&	$\tau_5$ \\
\hline
Magnitude	&	-0.30		&	0.50			&	-0.55		&	0.10			&	-0.45 \\
\hline
\end{tabular}
\caption{\label{tab:taus} Optimized values of \mbox{$\tau_1\dots\tau_5$} in units of inverse photon bandwidth.}
\end{table}

We apply the estimation procedure to the experimental \textsc{CNOT} gate and use $\mathscr{E}_\mathrm{max}$ and $\mathscr{E}_\mathrm{mean}$ to compare the gate model to the experimental gate. The results are summarized in Table \ref{tab:results}. It is evident that the optimized gate model agrees with the experimental data much better than the ideal gate model (\emph{i.e.} where mode-mismatch is ignored). The worst-case error observed is on the same order as recent maximum-likelihood QPT reconstructions performed on the same experimental gate (approximately 2.1\%) \cite{bib:Pryde04}. It should be noted that the model always predicts pure states in the extended Hilbert space and the $\tau$-parameters, which completely characterize the gate's operation, have a physical interpretation. When expectation values are calculated (\mbox{\emph{i.e.}} photo-detection is applied) these additional degrees of freedom are effectively traced out, which introduces mixture.
\begin{table}[!ht]
\begin{tabular}{|c|c|c|}
\hline
Gate model                           & $\mathscr{E}_\mathrm{max}$ & $\mathscr{E}_\mathrm{mean}$ \\
\hline
Ideal (\emph{i.e.} no mode-mismatch) & 15.25\%                    & 3.26\%                      \\
Optimized (globally)                 & 2.97\%                     & 1.32\%                      \\
Optimized (independently)            & 1.94\%                     & 0.67\%                      \\
\hline
\end{tabular}
\caption{\label{tab:results} Maximum and mean error between experimental \cite{bib:OBrien03} and predicted expectation values, using the ideal \textsc{CNOT} gate model, optimized gate model, and optimized gate model where parameters are estimated independently for each input setting.}
\end{table}

We perform QPT on the optimized gate model to construct a process matrix, which does not require a maximum-likelihood correction procedure since the model is inherently physical and self-consistent. From this we calculate the process fidelity with the ideal \textsc{CNOT} process, which yields \mbox{$F_P=0.88$}, consistent with the result produced through maximum-likelihood estimation of \mbox{$F_P=0.87$}. The process fidelity between the process matrices predicted by the two estimation procedures is \mbox{$F_P=0.95$}, indicating that the processes predicted by the two approaches are highly consistent. Thus, using a far smaller data set than is required for full QPT (64 \emph{vs.} 256 measurements) we predict essentially the same process matrix. This significant reduction comes about because we only need sufficient data to determine the five $\tau$-parameters from which any remaining data necessary for QPT can be inferred. This could be of particular importance in architectures where performing all of the measurements required for full QPT is prohibitive. Instead, the parameters describing the physical system could be determined from a smaller set of more accessible measurements, from which all of the data necessary for QPT can be inferred.

Next we repeat the estimation procedure where the mode-mismatch parameters are estimated independently for each input state used, allowing for input state dependent effects. This yields a significant improvement over the model with static parameters, as shown in Table \ref{tab:results}. This is indicative that input state dependent effects are a contributor to loss of gate fidelity. We infer that imperfections in the waveplates and their alignment change the mode-matching conditions in the circuit slightly for each different input state. Thus, by expanding the assumptions as to the physical nature of the processes taking place, one can significantly improve the accuracy of the model.

Experimentally, the target interferometer is assembled first and therefore ought to exhibit the least mode-mismatch. Thus, we expect the $\tau_4$ parameter to be smallest. This is consistent with our estimation of $\tau_4$. Following this, the control interferometer is assembled. Similarly, we expect the $\tau_1$ parameter to be the next smallest. Again, this is consistent with our results. Developing a complete understanding of circuit operation based on these figures, however, is complicated by the fact that the quoted parameters are globally optimized, whereas it is evident that input state dependent effects are significant. Thus, a complete interpretation would necessitate considering the parameters for the complete range of input states. Nevertheless, the parameters appear to give a consistent indication of the relative mode-match quality of different parts of the circuit.

The parameter $\tau_2$ only influences gate operation through interaction between the control and target qubits. In other words, in the absence of the control qubit, $\tau_2$ is \emph{invisible}. In general, parameters which do not affect interference will be invisible. For example, if we consider a displacement which affects all gate inputs, clearly this is non-interferometric and will not affect 2-qubit gate operation. However, if we were to embed the gate into say a 3-qubit circuit this parameter could well become interferometric. More generally, in order to fully characterize a given gate we may postulate that it be necessary to embed the gate into a higher dimensional circuit in order to observe all physical parameters. This is an important realization, since in reality we intend to operate gates as a part of larger circuits, not in isolation, and as such parameters which are invisible when operating in isolation could well become critical when operating in the context of a larger circuit \cite{bib:Oi03}.

We have presented a model for a \textsc{CNOT} gate in the LOQC architecture which explicitly allows for the effects of mode-mismatch. By fitting experimental data to this model we demonstrated that it is possible to infer the mode-matching characteristics of experimental gates, leading to significantly improved gate models. In the worst case, the error margin in the predictions made by the optimized gate models were shown to be similar to recent maximum-likelihood QPT studies (calculated using a smaller data set). Due to the counting statistics of the photon sources we expect gate error to exhibit a lower bound of approximately 1.5\% \footnote{Experimentally, single photons are derived from conditioned, spontaneous parametric down-conversion sources, which exhibit Poissonian counting statistics, the standard deviation of which is $c/\sqrt{c}$, where $c$ is the total number of measured counts. In our experiment we measured total counts of $c\approx 4600$, corresponding to $\sigma=1.47\%$.}. In the case of the optimized gate models the error margins were, on average, within this lower bound.

The ability to infer the physical characteristics of a gate represents a powerful diagnostic tool. It also allows us to perform QPT with a significantly reduced experimental data set, since only the parameters stipulated by the physical model need be determined, from which remaining measurement probabilities can be inferred. Because these probabilities are exact, the necessity for maximum-likelihood correction is mitigated. This approach to performing QPT differs from maximum-likelihood techniques in that it makes specific physical assumptions regarding the nature of the processes taking place and models these by expanding the Hilbert space to include non-qubit degrees of freedom. We also showed that by expanding the physical assumptions the model can be significantly improved. The model presented produces estimates which are inherently physical, pure and self-consistent, and relies on parameters with a physical interpretation. The techniques described are non-specific and could be applied to other optical circuits, or quantum computing architectures by first constructing suitable physical models.

We acknowledge Alexei Gilchrist, Andrew White and Austin Lund for helpful discussions. This work was supported by the Australian Research Council and ARDA under ARO contract number DAAD-19-01-1-0651.

\bibliography{paper.bib}

\end{document}